\documentstyle[12pt,fleqn]{article}
\def\beq{\begin{equation}}
\def\eeq{\end{equation}}
\def\5{\sigma_}
\def\6{\langle}
\def\9{\rangle}
\def\ot{\otimes}

\begin{document}

\renewcommand{\thefootnote}{\fnsymbol{footnote}}
\begin{center}
\vspace*{15mm}
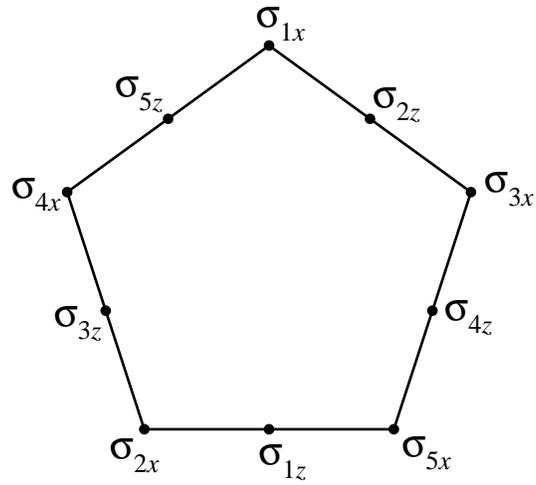
{\large{\bf Quantum codewords contradict local realism}}\\[10mm]

David P. DiVincenzo\medskip

{\it  IBM Research Division, IBM T. J. Watson Research
Center,\\ Yorktown Heights, New York, 10598}\\[5mm] and\\[5mm]

Asher Peres\footnote{Permanent address: Department of Physics,
Technion---Israel Institute of Technology, 32\,000 Haifa, Israel}

\medskip {\it Institute for Theoretical Physics, University
of California, Santa Barbara, CA 93106}\vfill

{\bf Abstract}\bigskip\end{center}

\begin{quote}
Quantum codewords are highly entangled combinations of two-state
systems. The standard assumptions of local realism lead to logical
contradictions similar to those found by Bell, Kochen and Specker,
Greenberger, Horne and Zeilinger, and Mermin. The new contradictions
have some noteworthy features that did not appear in the older
ones.\end{quote}\vfill


\vfill\newpage

Quantum codewords are highly entangled combinations of two-state quantum
sytems (qubits). They are structured in such a way that if one (or
sometimes more) of the qubits is perturbed, there remains enough
quantum information encoded in the remaining qubits for restoring the
original codeword unambiguously~\cite{Shor,Laflamme,Steane,Cald1}. In this
article, we shall investigate some properties of the 5-qubit codewords
invented by Bennett, DiVincenzo, Smolin, and Wootters~\cite{BDSW}
(which are equivalent, up to a change of bases of the individual 
qubits, to the five-qubit codewords of Laflamme 
{\em et al.}~\cite{Laflamme}).
The logical 0 is represented by the quantum state

\beq\begin{array}{l} |0_L\9=\mbox{$1\over4$}\,[-\,|00000\9\\
 \hspace*{17mm}-\,|11000\9-|01100\9-|00110\9-|00011\9-|10001\9\\
 \hspace*{17mm}+\,|10010\9+|10100\9+|01001\9+|01010\9+|00101\9\\
 \hspace*{17mm}+\,|11110\9+|11101\9+|11011\9+|10111\9+|01111\9],
\end{array} \label{codeword} \eeq
where, e.g., $|10010\9$ means $|1\9\ot|0\9\ot|0\9\ot|1\9\ot|0\9$, and
$|0\9$ and $|1\9$ are any two orthogonal states of a physical qubit. The
logical 1, denoted by $|1_L\9$, is obtained by exchanging all the
$|0\9$ and $|1\9$ in $|0_L\9$.  These two codewords have the useful
property of being invariant under cyclic permutations of the physical
qubits. This greatly simplifies the calculations below.

Let $\5x$, $\5y$, and $\5z$ be the standard Pauli spin matrices,
and $\5u$ denote the unit matrix (the latter will also be denoted by the
symbol 1, with no risk of error). It is convenient to introduce the
notation

\beq \5{abcde}\equiv
  \5{1a}\,\5{2b}\,\5{3c}\,\5{4d}\,\5{5e}\equiv
  \5a\ot\5b\ot\5c\ot\5d\ot\5e, \eeq
where the indices $abcde$ may be any combination of $u$, $x$, $y$, and
$z$. It is then readily verified that $|0_L\9$ and $|1_L\9$ are
eigenvectors, with eigenvalue 1, of the 32 following operators:
$\5{uuuuu}$, $\pm\5{zzzzz}$, and

\beq \5{xzuzx},\quad\5{yxuxy},\quad\5{zyuyz},\quad
 {\mp\5{uxzxu}},\quad{\mp\5{yuzuy}},\quad{\pm\5{xyzyx}}, \label{ops}
\eeq
and their cyclic permutations. The upper and lower signs refer to
$|0_L\9$ and $|1_L\9$, respectively (this convention will be followed
throughout this article). These 32 operators (with either choice
of sign) form an Abelian group; those whose sign does not change form an
invariant subgroup.  The existence of such a group associated with
this type of quantum error correction codes seems to be quite general.
A group-theoretic framework for codes
has been extensively developed by Gottesman~\cite{Got}
and by Calderbank {\em et al.}~\cite{Cald}.

It is well known that, for any entangled state, it is possible to find
operators whose correlations violate Bell's inequality
\cite{Capasso,GP}. However the codeword (\ref{codeword}) leads to a
stronger type of violation, without inequalities \cite{GHZ,Mermin}. 
In this article, it
will be shown that the codeword (\ref{codeword}) and its associated
operators (\ref{ops}) yield a rich crop of ``quantum paradoxes.''
It appears that these paradoxical properties are inherent to
all codewords of quantum error correcting codes. In
particular, this is obviously true
of the 9-qubit codewords of Shor \cite{Shor}, since the latter
are built from triads of Mermin states \cite{Mermin}.

It should be noted that the Mermin states,

\beq [|000\9\pm|111\9]/\sqrt{2}, \eeq
can be used as codewords, for correcting a ``bit error''
($0\leftrightarrow1$) in any one of the three qubits (but no other type
of error). These states are eigenvectors, with eigenvalue $+1$, of an
eight-element Abelian group:

\beq \5{uuu},\quad\mp\5{xyy},\quad\mp\5{yxy},\quad\mp\5{yyx},
  \quad\pm\5{xxx},\quad\5{zzu},\quad\5{zuz},\quad\5{uzz}. \eeq

To obtain quantum paradoxes for the five-qubit code (\ref{codeword}),
we note first that for each qubit, each one of $\5x$, $\5y$, and $\5z$
is an ``element of reality,'' as defined by Einstein, Podolsky, and Rosen
(EPR) \cite{EPR}. This is so because the observable value of any one of these
operators can be ascertained by measuring only {\it other\/} qubits,
``without disturbing in any way'' \cite{EPR} the element of reality
under consideration. For example, if we have prepared the five qubits
in the state $|0_L\9$, the result of a measurement of $\5{1x}$ can
be predicted with certainty by measuring $\5{2z}$ and $\5{3x}$, because
we know that $\5{1x}\5{2z}\5{3x}|0_L\9=-|0_L\9$.  Note that only the second
and third qubits have to be measured in order to determine $\5{1x}$
(it is not necessary to measure the fourth and fifth ones).  Other ways
of determining $\5{1x}$ without interacting with the first qubit are to
measure $\5{4x}\5{5z}$, or $\5{3x}\5{4y}\5{5y}$, or
$\5{2x}\5{3y}\5{4z}\5{5y}$, or $\5{2x}\5{3z}\5{5z}$, or
$\5{2y}\5{3y}\5{4x}$, or $\5{2y}\5{3z}\5{4y}\5{5x}$, or
$\5{2z}\5{4z}\5{5x}$, as may be seen from the various operators in
(\ref{ops}) and their cyclic permutations.

There are therefore eight different ways of determining $\5{1x}$ by
means of measurements performed on the {\it other\/} qubits. However,
these measurements cannot all be simultaneously carried out, if each
one of the qubits is tested separately, because they involve mutually
incompatible, non-commuting one-particle operators (although the eight {\it
products\/} of operators do commute, however, because their commutators
always involve an even number of anti\-commutations). The notion of
``element of reality'' tacitly implies that these eight different
determinations of $\5{1x}$ agree with each other. This may be
intuitively obvious.  However, classical intuition is a notoriously bad
guide in the quantum world.  There is no way of experimentally
verifying that the eight methods agree. (At most, it is possible to
verify that for some subsets of these operators, for example
$\5{2z}\5{3x}$ and $\5{4x}\5{5z}$ can be tested simultaneously. There
are only five such pairs among the eight operator products listed
above.) The assumption that all eight ways of determining $\5{1x}$
necessarily agree is manifestly counter\-factual. It is an example of
the metaphysical hypothesis known as {\it local realism\/}. This
hypothesis is incompatible with quantum mechanics, and leads to
numerous contradictions, as will now be shown.

As one example, among many, consider the following six operators:
$\pm\5{1z}\5{2z}\5{3z}\5{4z}\5{5z}$, and $\mp\5{1x}\5{2z}\5{3x}$ and
cyclic permutations of the five qubits. If we measure the values of
these six operators for one of the codewords, the result is 1, with
certainty. Actually, the easiest way of measuring any one of these
operators is to measure separately the physical qubits involved in it,
and then to multiply the results. It is therefore tempting to assume
that the values of the spin components of {\it individual\/} qubits
also satisfy

\beq v(\5{1z})\,v(\5{2z})\,v(\5{3z})\,v(\5{4z})\,v(\5{5z})=\pm1,
 \label{v5} \eeq
and

\beq v(\5{1x})\,v(\5{2z})\,v(\5{3x})=\mp1, \label{v3} \eeq
and all cyclic permutations of Eq.~(\theequation). There are six
equalities written above. The product of their right hand sides is
$-1$. But on the left hand side, each symbol appears twice, and
therefore the product of the left hand sides is $+1$. We have reached a
contradiction, of the same type as in refs.~\cite{GHZ} and
\cite{Mermin}. It is graphically illustrated in Fig.~1.

It is also possible to obtain a Bell-Kochen-Specker \cite{Bell66,KS}
type of contradiction, which does not refer to any particular quantum
state, such as (\ref{codeword}). Consider the following array of
operators:

$$\begin{array}{ccccccccccccc}
 \5{1z} & \5{2z} & \5{3z} & \5{4z} & \5{5z} & &1&1&1&1&1& \ &
 \5{1z}\5{2z}\5{3z}\5{4z}\5{5z}\\
 \5{1z} &1&1&1&1& &1& \5{2x} &1&1 & \5{5x} & & \5{5x}\5{1z}\5{2x}\\
 1& \5{2z} &1&1&1& & \5{1x} &1& \5{3x} &1&1& & \5{1x}\5{2z}\5{3x}\\
 1&1& \5{3z} &1&1& &1& \5{2x} &1& \5{4x} &1& & \5{2x}\5{3z}\5{4x}\\
 1&1&1& \5{4z} &1& &1&1& \5{3x} &1& \5{5x} & & \5{3x}\5{4z}\5{5x}\\
 1&1&1&1& \5{5z} &\quad& \5{1x} &1&1& \5{4x} &1& & \5{4x}\5{5z}\5{1x}
 \end{array} $$

\bigskip\noindent All the operators in that array have eigenvalues
$\pm1$, and therefore each one will yield one of these values, if
measured in the standard way.  Moreover, all the operators on each row
commute, and their product is 1.  Therefore, if all the operators on
one of the rows are actually measured, the product of the resulting
values is 1.  Likewise, all the operators in each column commute, and
their product is 1, {\it except those of the last column\/}, whose
product is $-1$. It is therefore clearly impossible to associate to
each operator a definite value, $\pm1$, that  is unknown but would be
revealed by a measurement of that operator, if such a measurement were
actually performed. This is the multiplicative form of the
Kochen-Specker contradiction \cite{qt,Mermin93}.

The original, additive form of the Kochen-Specker theorem can also be
obtained from the above array. In its original formulation, that theorem
asserted that there exist finite sets of projection operators,
such that it is impossible to attribute to each one
of the operators a bit value, ``true'' or ``false,'' subject to the
two following constraints:\\[3mm]
KS1) two orthogonal projection operators cannot both be true,
and\\[3mm]
KS2) if a subset of orthogonal projection operators is complete
(i.e., it has a sum equal to the unit operator), one of these
projection operators is true.\\[3mm]
In the physical interpretation of the Kochen-Specker theorem,
orthogonal projectors correspond to mutually compatible quantum
measurements, whose results are arbitrarily labelled 1 and 0, or
``yes'' and ``no.'' The theorem asserts that there exist sets of $n$
yes-no questions, such that none of the $2^n$ possible answers is
compatible with the sum rules of quantum mechanics. This implies that
there can be no subquantum physics, with hidden variables that would
ascribe definite outcomes to the $n$ yes-no tests (provided that the
hidden variables are not ``contextual,'' namely that the answer to each
question is unique, and does not depend on the choice of other
questions being asked).

A set of Kochen-Specker projectors can now be obtained from the above
array of operators as follows:

\medskip\noindent a) There is one complete set of eigenvectors that are
common to all the operators in the first row: it is the ``classical''
basis $|00000\9$, $|00001\9$, \ldots, $|11111\9$. The 32 projectors on
these vectors form a complete orthogonal set.

\medskip\noindent b) There is one complete set of eigenvectors that are
common to all the operators in the last column of the array. These are
the codewords $|0_L\9$ and $|1_L\9$, and the 15 mutations of each one
of them, obtained by letting one of the Pauli matrices act on one of
the physical qubits. The 32 projectors on these orthonormal vectors
form another complete set. Each one is moreover orthogonal to 16
vectors of the ``classical'' basis, and vice-versa.

\medskip\noindent c) Each one of the five other rows generates eight
mutually orthogonal 4-dimensional subspaces, that form a complete set.
For example, the subspaces that correspond to the third row are the
tensor products of the eigenvectors of $\5{1x}$, $\5{2z}$,
$\5{3x}$, and the complete subspaces of the two other qubits. The
products of the three eigenvectors are

\beq \mbox{$1\over2$}\,(|0\9\pm|1\9)
 \ot(|0\9\;{\rm or}\;|1\9)\ot(|0\9\pm|1\9), \eeq
or

\beq\begin{array}{lll}
 \mbox{$1\over2$}\,(|000\9+n\,|001\9+m\,|100\9+mn\,|101\9) &
 {\rm for} & \5{2z}>0,\\
 \mbox{$1\over2$}\,(|010\9+n\,|011\9+m\,|110\9+mn\,|111\9) &
 {\rm for} & \5{2z}<0,\end{array}\eeq
where $m=\6\5{1x}\9$ and $n=\6\5{3x}\9$.
The eight corresponding projection operators thus are

\beq 
 \mbox{$1\over4$}(|000\9+n|001\9+m|100\9+mn|101\9)
 (\6000|+n\6001|+m\6100|+mn\6101|)\!\ot1\!\ot1,\;\eeq
and

\beq 
 \mbox{$1\over4$}(|010\9+n|011\9+m|110\9+mn|111\9)
 (\6010|+n\6011|+m\6110|+mn\6111|)\!\ot1\!\ot1,\;\eeq
respectively. There are therefore 40 projectors of rank 4. They satisfy
many mutual ortho\-gonality relations, for example, any projector with
$\6\5{1x}\9=1$ in the third row of the array is orthogonal to any
projector with $\6\5{1x}\9=-1$ in the sixth row.

Moreover, any rank~4 projector is orthogonal to many of the 64
projectors of rank~1, listed above. For example, all the projectors in
(\theequation), for any $m$ and $n$, are orthogonal to all the
``classical'' vectors $|a0cde\9$. All the projectors in (\theequation)
with $m=n$ (so that $\6\5{1x}\5{2z}\5{3x}\9=-1$) are orthogonal to
$|1_L\9$ and to all its mutations of type $\5{4d}\5{5e}|1_L\9$, and to
some others. They are also orthogonal to the various mutations of
$|0_L\9$, generated by $\5{1y}$, $\5{1z}$, $\5{2x}$, $\5{2y}$,
$\5{3y}$, $\5{3z}$, or any odd number of the latter.  (Not all these
vectors are distinct, however.)

These numerous orthogonality relations have as a consequence that the
constraints KS1 and KS2 cannot both be satisfied. The novel features in
this Kochen-Specker contra\-diction is that projectors of rank 4 are
used, and that the total number of projectors involved is remarkably
low, when compared to the number of dimensions:  $104/32=3.25$, while a
similar construction in 4 dimensions requires 24 vectors~\cite{JPA}, and
in 8 dimensions, 40 vectors are involved~\cite{KP}.

We have likewise investigated the 7-qubit codewords of
Steane~\cite{Steane}. They are simultaneous eigenvectors of 128
matrices of order 128, which are direct products of 3 to 7 Pauli
matrices, and form an Abelian group. There are subsets of 10 group
elements with properties similar to those listed in (\ref{v5}) and
(\ref{v3}): each Pauli matrix corresponds to a local ``element of
reality,'' because the result of its measurement can be predicted with
certainty by examining only {\it other\/} qubits. However, if it is
assumed, in accordance with local realism, that each one of the local
Pauli matrices is associated with a definite numerical value, $\pm1$,
an algebraic contradiction appears.

\bigskip 
We are grateful for the hospitality of the Program on Quantum Computers
and Quantum Coherence at the Institute for Theoretical Physics,
University of California at Santa Barbara, where this work was completed
and was supported in part by the National Science
Foundation under Grant No.\ PHY94-07194.
DPDV acknowledges support from the Program on the Comparative Analysis
of Quantum Computer Implementations of the U.~S. Army Research Office.

\vfill

\noindent FIG. 1. \ Each side of the pentagon corresponds to three
mutually compatible measurements. The product of the three results is
guaranteed to have value $\mp1$, for $|0_L\9$ and $|1_L\9$, respectively.
Moreover, the product of the five $\5z$ has to be $\pm1$. There is no
consistent set of values for the twelve operators.

\clearpage

\begin{thebibliography}{99}

\bibitem{Shor} P. W. Shor, Phys. Rev. A {\bf 52}, 2493 (1995).
\bibitem{Laflamme} R. Laflamme, C. Miquel, J. P. Paz, and W. H. Zurek,
Phys. Rev. Lett. {\bf 77}, 198 (1996).
\bibitem{Steane} A. M. Steane, Phys. Rev. Lett. {\bf 77}, 793 (1996);
Proc. Roy. Soc. (London) A {\bf 452}, 2551 (1996).
\bibitem{Cald1} A.~R. Calderbank and P.~W. Shor, Phys. Rev. A {\bf 54},
1098 (1996).
\bibitem{BDSW} C. H. Bennett, D. P. DiVincenzo, J. A. Smolin,
and W. K. Wootters, Phys. Rev. A, {\bf 54}, 3824 (1996).
\bibitem{Got} D. Gottesman, Phys. Rev. A {\bf 54}, 1862 (1996).
\bibitem{Cald} A.~R. Calderbank, E.~M. Rains, P.~W. Shor, and
N.~J.~A. Sloane, ``Quantum Error Correction and Orthogonal Geometry,''
e-print quant-ph/9605005; ``Quantum Error Correction via Codes over
GF(4),'' e-print quant-ph/9608006, submitted to IEEE Transactions on
Information Theory.
\bibitem{Capasso} V. Capasso, D. Fortunato, and F. Selleri, Int. J.
Theor. Phys. {\bf 7}, 319 (1973).
\bibitem{GP} N. Gisin and A. Peres, Physics Letters A {\bf 162}, 15 (1992).
\bibitem{GHZ} D. M. Greenberger, M. Horne, and A. Zeilinger, in
{\it Bell's Theorem, Quantum Theory, and Conceptions of the Universe\/}
ed. by M. Kafatos (Kluwer, Dordrecht, 1989) p.~69.
\bibitem{Mermin} N. D. Mermin, Physics Today {\bf 43} (no. 6), 9 (1990);
Am. J. Phys. {\bf 58}, 731 (1990).
\bibitem{EPR} A. Einstein, B. Podolsky, and N. Rosen, Phys. Rev. {\bf 47},
777 (1935). 
\bibitem{Bell66} J. S. Bell, Rev. Mod. Phys. {\bf 38}, 447 (1966).
\bibitem{KS} S. Kochen and E. P. Specker, J. Math. Mech. {\bf 17}, 59
(1967).
\bibitem{qt} A. Peres, {\it Quantum Theory: Concepts and Methods\/}
(Kluwer, Dordrecht, 1993) p.~189.
\bibitem{Mermin93} N. D. Mermin, Rev. Mod. Phys. {\bf 65}, 803 (1993).
\bibitem{JPA} A. Peres, J. Phys. A {\bf 24}, L175 (1991).
\bibitem{KP} M. Kernaghan and A. Peres, Physics Letters A {\bf 198}, 1
(1995).  

\end{thebibliography}
\end{document}